# SLI, a New Metric to determine Success of a Software Project

Shashi Kumar N.R.
Advanced Software Engineering
Research Group, RIIC
Dayananda Sagar Institutions
Bangalore, India
nrshash@gmail.com

T.R. Gopalakrishnan Nair
Advanced Software Engineering
Research Group, RIIC, DSI
Aramco Endowed Chair -
Technology, PMU, KSA.
Bangalore, India
trgnair@ieee.org

Suma V
Advanced Software Engineering
Research Group, RIIC
Dayananda Sagar Institutions
Bangalore, India
sumavdsce@gmail.com

**Abstract-** Project Management process plays a critical role in managing factors such as cost, time, technology and personnel towards achieving the success of a project and henceforth the sustainability of the company in the industrial market. This paper emphasizes empirical study of several projects developed over a period of time in a product and service based CMMI Level 5 Software Company. The investigation shows impact analysis of resources such as cost, time, and number of developers towards the successful completion of the project as allocated by the project manager during the developmental process. The analysis has further led to the introduction of a new qualitative metric, Success Level Index Metric (SLI) whose index value varies from 0 to 1. SLI acts as a maturity indicator that indicates the degree of maturity of the company in terms of success of their projects based on which the company can choose their desired level of success for their projects.

## I. INTRODUCTION

Software, due to its varied significance and diversity, a high quality of software is expected to be developed while considering the optimization of time and cost. Developing quality software requires a challenging effort as it involves complex process. Customer satisfaction is one of the paramount issue in engineering a high quality artefacts.

Hence, a project management process has a role to control measure and improve the quality of product through the developmental process [1]. Project management process is therefore justifiable for the successful initiation, planning, design, execution, monitoring, controlling and in successful closure of the project. Subsequently, it is vital for the project management process to establish project objectives and its expected success criteria for determining the right choice of resources at the right time to accomplish the formulated tasks in the project. However, success criteria is not evaluated based upon the successful delivering of project within estimated cost, time but is also decided on the level of quality in the product that satisfies customer requirement and which enhances the profit of the company [2]. Hence, a comprehensive analysis of efficiency of project management process in accurate estimation and allocation of optimal resources towards the development of projects decides the success level of the project and henceforth the sustainability of the company.

Authors in [3] expresses that in any organization Continuous improvement (CI) is an ongoing activity and through focused incremental changes in process, organizational performance can be raised. Authors in[4] inform that a continuous improvement in process provides a planned and organized system for the continual discovery and implementation of such process changes.

The main objective of this paper is to analyze the impact of critical resources such as cost, time and number of developers towards the successful completion of the project as allocated during the developmental process. The organization of the paper is as follows. Section II of the paper briefs about the background work for this investigation. Section III presents the research Methodology, Section IV provides the empirical analysis of several projects through a case study and Section V summarizes the paper.

## II. RELATED WORK

Authors in [4] suggests that project manager is highly responsible in balancing and satisfying the challenging demands in a project due to scope, time, cost, risk etc. He further states that quality of the product depends on the aforementioned parameters. Authors in [5] feels that project planning is one of the most important activities in any software project. Hence, they recommend sensitivity of project managers in increasing the success level of the projects.

Authors in [6] discussing management issues, project managers often classify projects as either simplex or complex which indicates complexity makes a difference to the project management. Authors propose a worker's skill-based metric for the complexity of a construction project activity using the concepts of functional decomposition and validate it. Authors in this paper identify the influence of the skill of the workers on the complexity of the project activity and their support to project manager's understanding of the complexity. Hence, research made by authors in [7] indicates there is a linkage between the personality and project manager performance.

Authors in [8] explains to enhance the efficiency and effectiveness with which Six Sigma projects are initiated, planned, executed and closed. They introduced a new model for the integration and application of Six Sigma and project management methods.

Authors in [9] suggests that the completion of a project requires input from a variety of groups including client, project team, parent organization, the producer and the end user. Every member has a role in defining and determining the success. They all have specific roles and responsibilities to attain success. Authors in [10] claims that People produce success. People are a project manager's most valuable asset, tools and processes are unproductive without good people. Thus, project leader better describe what project managers must be to produce success on projects.

Authors in [11] discussing management issues, project managers often classify projects as either simplex or complex which indicates complexity makes a difference to the project management. Authors propose a worker's skill-based metric for the complexity of a construction project activity using the concepts of functional decomposition and validate it. Authors in this paper identify the influence of the skill of the workers on the complexity of the project activity and their support to project manager's understanding of the complexity.

Authors in [12] say that large projects will have indefinite durations with more number of tasks. This results no well-defined critical paths. It is beneficial to position time buffers to help ensure the completion of subtasks and the overall project but because of the complexity it is difficult to determine the best solution. The best plans are likely to change over time as uncertainties; therefore they have introduced a new approach that can be used to position time buffers at each project review.

Authors in [13] say that the Information Technology investments have long durations and require significant amount of capital expenditure. Traditional tools are proven insufficient to identify the true values. Most IT managers do not have the necessary skills to carry ROA (Real Option Analysis]. In this paper, they propose an easy-to-use framework for using ROA to value investment opportunities.

Authors in [14] examine the engineering–procurement–construction (EPC) project to estimate an optimal schedule. Considering the precedence and limited resource constraints minimum make span of the project is determined.

Authors in [15] say expertise from various functional departments is required for complex projects with large number of tasks. During the planning stage of a project, the assignment of people to teams and tasks is an important step. The lack of communication and cooperation among team members could seriously delay the project completion if a project does not have effective teams to work on it. Team member characteristics such as multifunctional knowledge, teamwork capabilities and working relationship with quantifiable measures and each member's workload schedule are used in the proposed model for task-member assignments, so that the right team member will be selected for the right task at the right time.

Authors in [16] says that large projects will have indefinite durations with more number of tasks. This results no well-defined critical paths. It is beneficial to position time buffers to help ensure the completion of subtasks and the overall project but because of the complexity it is difficult to determine the best solution. The best plans are likely to change over time as uncertainties, therefore they have introduced a new approach that can be used to position time buffers at each project review.

Authors in [17] says that the Information Technology investments have long durations and require significant amount of capital expenditure. Traditional tools are proven insufficient to identify the true values. Most IT managers do not have the necessary skills to carry ROA (Real Option Analysis]. In this paper, they propose an easy-to-use framework for using ROA to value investment opportunities.

Authors in [18] says expertise from various functional departments is required for complex projects with large number of tasks. During the planning stage of a project, the assignment of people to teams and tasks is an important step. The lack of communication and cooperation among team members could seriously delay the project completion if a project does not have effective teams to work on it. Team member characteristics such as multifunctional knowledge, teamwork capabilities and working relationship with quantifiable measures and each member's workload schedule are used in the proposed model for task-member assignments, so that the right team member will be selected for the right task at the right time.

This investigation therefore aims to study the impact of project influencing parameters such as cost, time, and number of developers towards the realization of successful projects.

## III. RESEARCH METHODOLOGY

The aim of this research is to analyse the success level of a project based on various critical resources. Hence, this investigation comprises of a case study made in a leading service based software industry having CMMI Level 5 and ISO certifications. Several projects were investigated in order to analyze the impact of various resources towards the success of the project during software development process. The data was collected for the purpose of analysis from quality assurance department and document management repository of the company under study. Analysis of data infers that efficiency of project management process has a greater impact towards success of the project.

## IV. CASE STUDY

CMMI Level 5 and ISO certified service/product based industry was considered for case study. The company operates their business on Business Intelligence, data warehouse etc.

Table 1 depicts a sampled data of twenty five projects developed since 2009 to 2012. These projects are developed in object supporting technology using languages such as .net and Java. Table 1 provides the resource information as estimated and actual resource information as occurred during developmental process in addition to the variation observed between the values. The resource information projected in the table includes cost, time, and number of developers assigned, number of defects in addition to the success level of the complete project.

TABLE 1: SUCCESS LEVEL OF A PROJECT WITH RESPECT TO VARIATIONS IN RESOURCES

| PROJECT | NO. OF DEVELOPERS | | | DEFECTS | | | TIME | | | COST($) | | | Success Level of a Project |
|---|---|---|---|---|---|---|---|---|---|---|---|---|---|
| | EST. | ACT. | % | EST. | ACT. | % | EST. | ACT. | % | EST. | ACT. | % | |
| P1 | 1 | 1 | 0 | 36 | 30 | 15.6 | 132 | 128 | 2.9 | 3296 | 3200 | 2.9 | 94.64 |
| P2 | 1 | 2 | 100 | 48 | 45 | 6.9 | 218 | 174 | 20 | 5438 | 4350 | 20 | 63.28 |
| P3 | 1 | 1 | 0 | 64 | 67 | -4 | 278 | 232 | 16.7 | 6960 | 5800 | 16.7 | 92.66 |
| P4 | 1 | 1 | 0 | 67 | 65 | 2.5 | 288 | 240 | 16.7 | 7200 | 6000 | 16.7 | 91.04 |
| P5 | 1 | 1 | 0 | 69 | 72 | -4.5 | 298 | 248 | 16.7 | 7440 | 6200 | 16.7 | 92.8 |
| P6 | 1 | 2 | 100 | 100 | 95 | 5 | 432 | 360 | 16.7 | 10800 | 9000 | 16.7 | 65.42 |
| P7 | 1 | 1 | 0 | 144 | 145 | -0.4 | 582 | 520 | 10.7 | 14560 | 13000 | 10.7 | 94.74 |
| P8 | 1 | 2 | 100 | 162 | 159 | 2 | 701 | 584 | 16.7 | 17520 | 14600 | 16.7 | 66.17 |
| P9 | 1 | 2 | 100 | 200 | 201 | -0.5 | 907 | 720 | 20.6 | 22680 | 18000 | 20.6 | 64.81 |
| P10 | 1 | 2 | 100 | 200 | 198 | 1 | 792 | 720 | 9.1 | 19800 | 18000 | 9.1 | 70.2 |
| P11 | 1 | 2 | 100 | 211 | 215 | -1.8 | 836 | 760 | 9.1 | 20900 | 19000 | 9.1 | 70.92 |
| P12 | 1 | 2 | 100 | 211 | 215 | -1.8 | 912 | 760 | 16.7 | 22800 | 19000 | 16.7 | 67.13 |
| P13 | 1 | 2 | 100 | 356 | 349 | 1.8 | 1472 | 1280 | 13 | 36800 | 32000 | 13 | 68.02 |
| P14 | 1 | 2 | 100 | 367 | 353 | 3.7 | 1478 | 1320 | 10.7 | 36960 | 33000 | 10.7 | 68.71 |
| P15 | 2 | 3 | 50 | 456 | 459 | -0.8 | 2017 | 1640 | 18.7 | 50430 | 41000 | 18.7 | 78.34 |
| P16 | 2 | 2 | 0 | 478 | 481 | -0.7 | 2133 | 1720 | 19.4 | 53320 | 43000 | 19.4 | 90.49 |
| P17 | 2 | 3 | 50 | 536 | 535 | 0.1 | 2314 | 1928 | 16.7 | 57840 | 48200 | 16.7 | 79.14 |
| P18 | 2 | 2 | 0 | 589 | 595 | -1 | 2332 | 2120 | 9.1 | 58300 | 53000 | 9.1 | 95.71 |
| P19 | 3 | 5 | 66.7 | 1011 | 1100 | -8.8 | 4404 | 3640 | 17.4 | 110110 | 91000 | 17.4 | 76.85 |
| P20 | 3 | 5 | 66.7 | 1111 | 1200 | -8 | 4200 | 4000 | 4.8 | 105000 | 100000 | 4.8 | 82.95 |
| P21 | 3 | 3 | 0 | 1133 | 1120 | 1.2 | 4570 | 4080 | 10.7 | 114240 | 102000 | 10.7 | 94.35 |
| P22 | 5 | 7 | 40 | 1700 | 1665 | 2.1 | 7344 | 6120 | 16.7 | 183600 | 153000 | 16.7 | 81.15 |
| P23 | 5 | 7 | 40 | 1789 | 1652 | 7.7 | 7599 | 6440 | 15.3 | 189980 | 161000 | 15.3 | 80.46 |
| P24 | 6 | 8 | 33.3 | 1944 | 1532 | 21.2 | 7700 | 7000 | 9.1 | 192500 | 175000 | 9.1 | 81.82 |
| P25 | 6 | 7 | 16.7 | 1989 | 1851 | 6.9 | 8592 | 7160 | 16.7 | 214800 | 179000 | 16.7 | 85.77 |

P1...P25 – Projects; Est - Estimated, ACT - Actual

This research led towards the introduction of new metric namely Success Level Index (SLI). SLI is a quality measurement metric aimed towards measuring the success level of the projects.

Success Level Index (SLI) = (Achieved Success Level in the Project)/ (Expected Success Level in the Project)

(1)

Further, it is worth to note that the sampled projects depicted in this paper are business application projects and hence SLI is evaluated in compliance with SLI of non-critical applications. Table 2.1therefore illustrates SLI for the non-critical applications and Table 2.2 represents SLI for critical applications project whose values vary from 0 to 1.

TABLE 2.1: SLI, SERVICE LEVEL INDEX FOR NON-CRITICAL APPLICATIONS

| SLI - Success Level Index (Non-Critical applications) | |
|---|---|
| 0.0 – < 0.5 | Complete Failure (CF) |
| 0.5 – < 0.7 | Product Not Intended Level (PNI) |
| 0.7 – < 0.8 | Possible Rework Level (PR) |
| 0.8 – < 0.9 | Needs Improvement Level (NI) |
| 0.9 – 1 | Ideal Level (I) |

TABLE 2.2: SLI, SERVICE LEVEL INDEX FOR CRITICAL APPLICATIONS

| SLI - Success Level Index (Critical applications) | |
|---|---|
| 0 –< 0.6 | Complete Failure (CF) |
| 0.6 –< 0.8 | Product Not Intended Level (PNI) |
| 0.75 –<0.85 | Possible Rework Level (PR) |
| 0.85 –< 0.95 | Needs Improvement Level (NI) |
| 0.95 – 1 | Ideal Level (I) |

Benefits of SLI
- SLI is a quality indicator metric which indicates the quality level of the project in terms of success
- SLI acts as an maturity indicator metric that indicates the degree of maturity of the company in terms of success of their projects
- SLI is a decisive metric upon which the company can choose their desired level of success for their projects
- With the introduction of SLI, this research further directed towards analysis of the sampled projects in terms of SLI.

According to the Software Quality Assessment (SQA) policy of the company under study, it is stringent for every project to be developed within 10 percent variation of estimation and actual observation. Any divergence from the specified percent is deemed to be threat to the success of the project and henceforth the sustainability of the company in the industrial market.

TABLE 3: VARIATION AND SLI

| PROJECTS | No of Developers | Defects | Time/Cost | Success Level of a Project (%) |
|---|---|---|---|---|
| P2 | H | M | H | 63.28 |
| P9 | H | M | H | 64.81 |
| P6 | H | M | H | 65.42 |
| P8 | H | M | H | 66.17 |
| P12 | H | M | H | 67.13 |
| P13 | H | M | M | 68.02 |
| P14 | H | M | M | 68.71 |
| P10 | H | M | M | 70.20 |
| P11 | H | M | M | 70.92 |
| P19 | H | M | H | 76.85 |
| P15 | H | M | H | 78.34 |
| P17 | H | M | H | 79.14 |
| P23 | H | M | H | 80.46 |
| P22 | H | M | H | 81.15 |
| P24 | H | H | M | 81.82 |
| P20 | H | M | M | 82.95 |
| P25 | H | M | H | 85.77 |
| P16 | N | M | H | 90.49 |
| P4 | N | M | H | 91.04 |
| P3 | N | M | H | 92.66 |
| P5 | N | M | H | 92.80 |
| P21 | N | M | M | 94.35 |
| P1 | N | M | M | 94.64 |
| P7 | N | M | M | 94.74 |
| P18 | N | M | M | 95.71 |

| N | NO VARIATION | 0% |
|---|---|---|
| M | MODERATE VARIATION | < 15% |
| H | HIGH VARIATION | > 15% |

Table 3 illustrates the variations observed in the resource values as estimated by the project manager and the actual resources subsequently assigned by the same project manager for the project completion. In compliance with the organizational SQA policy, a deviation in the resources greater than 10% from the estimation is deemed to be High (H) and less than 10% is considered to be moderate (M) in variations. If the estimated and observed resources remain unchanged, the project is considered to be developed with no variations (N) in the resources. It is worth to note that the table comprises of same sampled projects as depicted in Table 1. However, they are arranged in the ascending order of SLI.

CF level indicates that product which is developed is a complete failure and hence to be discarded as it does not yield customer satisfaction at any level. PNI status of a project indicates that the product is not the right product that should have been developed. Hence, it also does not satisfy the customer. A project which is at PR level demands reworking on the product right from the SRS onwards till its implementation. However, when a project is developed having a status of NI indicates that few functionalities, which are not up to the expectations of the customer alone requires improvement in order to achieve total customer satisfaction. A project is deemed to be ideal in terms of success level if satisfies the customer in total. It is worth to note that SLI varies from critical applications which include machine critical and flight critical applications and non-critical applications which includes business applications.

## V. CONCLUSION

The observational inferences drawn in the paper indicate the need to evaluate the impact of cost, time, and number of developers parameters analytically on the success level. The introduction of a Metric SLI, Service Level Index of a project further would help the PMs to classify the success level of software project. As a result of this work, it is imperative for the companies to dedicate much more attention while allocating the resources and to minimize the differences in variations in estimation and actual resources allocation. Therefore an utmost importance in this area is demanded and expected from a Project management process lest affects the quality and success of a project and a company.

ACKNOWLEDGMENT

The authors would like to sincerely acknowledge all the industry people for their immense help in carrying out this work. The authors would like to thank all the referees and the editor for their constructive and helpful comments.